\documentclass[revtex4,apj,iop]{emulateapj}
\usepackage{epsfig}
\usepackage{amsmath}
\usepackage{natbib}
\usepackage{graphicx}
\usepackage{changes}

\def \TI {$^{44}$Ti}

\newcommand{\NS}{\mbox {\it NuSTAR}}
\newcommand{\XMM}{\mbox {\it XMM-Newton}}
\newcommand{\SUZ}{\mbox {\it Suzaku}}

\newcommand{\CHAN}{\mbox {\it Chandra}}

\begin{document}

\title{Locating the most energetic electrons in Cassiopeia A}

\author{
Brian~W.~Grefenstette\altaffilmark{1},
Stephen~P~Reynolds\altaffilmark{2},
Fiona~A.~Harrison\altaffilmark{1},
T.~Brian~Humensky\altaffilmark{3},
Steven~E.~Boggs\altaffilmark{4},
Chris~L.~Fryer\altaffilmark{5},
Tracey~DeLaney\altaffilmark{6},
Kristin~K.~Madsen\altaffilmark{1},
Hiromasa~Miyasaka\altaffilmark{1},
Daniel~R.~Wik\altaffilmark{7, 8},
Andreas~Zoglauer\altaffilmark{4},
Karl~Forster\altaffilmark{1},
Takao~Kitaguchi\altaffilmark{9},
Laura~Lopez\altaffilmark{10},
Melania~Nynka\altaffilmark{11},
Finn E. Christensen\altaffilmark{12},
William W. Craig\altaffilmark{4},
Charles J. Hailey\altaffilmark{11},
Daniel Stern\altaffilmark{13},
William W. Zhang\altaffilmark{6}
}

\def \cahill {Cahill Center for Astrophysics, 1216 E. California Blvd, California Institute of Technology, Pasadena, CA 91125, USA}
\altaffiltext{1}{\cahill}
\def \ncstate {Physics Department, NC State University, Raleigh, NC 27695, USA}
\altaffiltext{2}{\ncstate}
\altaffiltext{3}{Physics Department, Columbia University, New York, NY 10027, USA}
\altaffiltext{4}{Space Sciences Laboratory, University of California, Berkeley, CA 94720, USA}
\altaffiltext{5}{CCS-2, Los Alamos National Laboratory, Los Alamos, NM 87545, USA}
\altaffiltext{6}{Physics \& Engineering Department, West Virginia Wesleyan College, Buckhannon, WV 26201, USA}
\altaffiltext{7}{Astrophysics Science Division, NASA/Goddard Space Flight Center, Greenbelt, MD 20771, USA}
\altaffiltext{8}{Department of Physics and Astronomy, Johns Hopkins University, Baltimore, MD 21218, USA}
\altaffiltext{9}{RIKEN, 2-1 Hirosawa, Wako, Saitama, 351-0198, Japan}
\altaffiltext{10}{MIT-Kavli Institute for Astrophysics and Space Research, 77 Massachusetts Avenue, 37-664H, Cambridge, MA 02139, USA }
\altaffiltext{11}{Columbia Astrophysics Laboratory, Columbia University, New York, NY 10027, USA }
\altaffiltext{12}{DTU Space, National Space Institute, Technical University of Denmark, Elektrovej 327, DK-2800 Lyngby, Denmark}
\altaffiltext{13}{Jet Propulsion Laboratory, California Institute of Technology, Pasadena, CA 91109, USA}

\email{bwgref@srl.caltech.edu}



%
%

\begin{abstract}

We present deep ($>$2.4 Ms) observations of the Cassiopeia A supernova remnant with {\it NuSTAR}, which operates in the 3--79 keV bandpass and is the first instrument capable of spatially resolving the remnant above 15 keV. We find that the emission is not entirely dominated by the forward shock nor by a smooth ``bright ring" at the reverse shock. Instead we find that the $>$15 keV emission is dominated by knots near the center of the remnant and dimmer filaments near the remnant's outer rim. These regions are fit with unbroken power-laws in the 15--50 keV bandpass, though the central knots have a steeper ($\Gamma \sim -3.35$) spectrum than the outer filaments ($\Gamma \sim -3.06$). We argue this difference implies that the central knots are located in the 3-D interior of the remnant rather than at the outer rim of the remnant and seen in the center due to projection effects. The morphology of $>$15 keV emission does not follow that of the radio emission nor that of the low energy ($<$12 keV) X-rays, leaving the origin of the $>$15 keV emission as an open mystery. Even at the forward shock front we find less steepening of the spectrum than expected from an exponentially cut off electron distribution with a single cutoff energy. Finally, we find that the GeV emission is not associated with the bright features in the {\it NuSTAR} band while the TeV emission may be, suggesting that both hadronic and leptonic emission mechanisms may be at work.

\end{abstract}

\section{Introduction}
\label{section:intro}

Young supernova remnants such as Cassiopeia A (Cas A) with shock velocities
above 1000 km s$^{-1}$ provide excellent opportunities to study in
detail the process of shock acceleration of electrons to
high energies (see \citealt{Reynolds:2008fia} for a review).  As the youngest
Galactic remnant of a core-collapse (CC) supernova with an estimated explosion
date of 1670 \citep{Thorstensen:2001dza}, Cas A 
particularly important for contrasting with the historical remnants of
Type Ia events, such as Tycho, Kepler, SN 1006, and G1.9+0.3 (although the identification
as a Type Ia is less secure for G1.9+0.3). Non-thermal X-ray emission from all these
objects can be characterized both spectrally and spatially and can be used to
infer properties of the acceleration process. Of special interest for shock acceleration
physics is the maximum energy to which electrons are accelerated, $E_m$, and its dependence
on the local shock velocity and other parameters. For example, the ``thin rims'' of synchrotron
X-rays found at the peripheries of some supernova remnants imply strong
magnetic-field amplification \citep{2003ApJ...584..758V, Parizot:2006dz, 2014ApJ...790...85R}
which, along with spectral inferences, gives an assumed exponential cutoff
in the electron spectrum with $e$-folding energies in the range 10 -- 100 TeV.

Cas A has now been shown by light echoes to be the result of a SN IIb explosion where the progenitor lost
most of a massive H envelope prior to the explosion \citep{Krause:2008bda}. This implies that the remnant is
currently expanding into the progenitor's stellar wind and so the blast wave should be quasi-perpendicular
over most of its surface. Since the properties of shock acceleration are strongly dependent on the magnetic obliquity angle $\theta_{\rm Bn}$
between the shock velocity and the upstream magnetic field, one expects substantial differences in both the morphology and spectrum of synchrotron emission between
remnants of CC and Type Ia events. In particular, many models (e.g. \citealt{2010ApJ...717.1054R}) assert that acceleration at
quasi-perpendicular shocks ($\theta_{\rm Bn} \sim 90^\circ$) is considerably different from that at quasi-parallel shocks, so that
remnants expanding into stellar winds might have very different properties in their non-thermal emission.

 
The spectral properties and basic morphology of Cas A in X-rays
between 0.5 and 12 keV are well known from many previous studies,
especially with {\it XMM-Newton} \ \citep{2001A&A...365L.225B, 2002A&A...381.1039W}
and {\it Chandra} \ \citep{Gotthelf:2001bn, Hwang:2004dua}. 
Thermal emission from the shocked ejecta dominates the integrated spectrum at these energies,
with electron temperatures typically between 1 and 3 keV and
in no cases above 6 keV. Local variations have been mapped out on
arcsecond length scales \citep{Hwang:2012gi}; in almost all regions, a
power-law component must be added to one or more thermal components to
obtain a satisfactory description of the spectrum.

The morphology is dominated by a ``bright ring" about 3.5$^{\prime}$ in diameter which is
commonly associated with the reverse shock or contact discontinuity between shocked ejecta and
shocked circumstellar material (CSM). Fainter emission forms a faint
rim outside the bright ring \citep{Gotthelf:2001bn}. The bright ring
remains prominent to the highest energies imaged with {\it XMM-Newton} \
\citep{2001A&A...365L.225B}, with an East/West asymmetry reported in observed using {\it Beppo-Sax} \citep{Vink:2000bu}
and~\SUZ~\citep{Maeda:2009vla}. These authors note that the western half of the remnant appears to be brighter
in the highest energy band of both instruments. The faint rim is located at the edge
of radio emission, though the outer radio emission appears to form the edge of a plateau rather than thin filaments.
Many arguments suggest that the outer radio edge and the X-ray filaments are located at the outer blast wave.

The integrated spectrum of Cas A has long been known to continue above
the thermal spectrum to energies of order 100 keV with a
spectrum reasonably well described by a single power-law, based on
data from non-imaging instruments (e.g. {\it CGRO}, \citealt{1996AAS..120C.357T}; {\it RXTE},
{\it HEAO-2}, and {\it OSSE}, \citealt{1997ApJ...487L..97A}; {\it Beppo-SAX}, \citealt{Vink:2000bu}; {\it INTEGRAL}, \citealt{Renaud:2006gra}; \SUZ, \citealt{Maeda:2009vla}). 
While the tail was originally thought to be thermal, at least out to energies of
30 keV or so \citep{Pravdo:1979ik}, better analysis of the spectrum
below 10 keV has shown that the temperature of the thermal plasmas only extend to $\sim$3 keV \citep{Hwang:2012gi}.
Subsequent explanations on the high-energy tail have included non-thermal bremsstrahlung from a
power-law distribution of somewhat suprathermal electrons (e.g. \citealt{Asvarov1990}, 
\citealt{2001A&A...365L.225B}) and synchrotron emission
from shock-accelerated electrons with much higher energies (e.g.,
\citealt{1997ApJ...487L..97A}). While the hard tail can be associated with the
power-law components required in the modeling of the emission below 10 keV,
the temperature and flux of the thermal components vary with position in the remnant. It
is not clear how much of the flux below 10 keV is associated with the non-thermal
component. In relatively line-free regions of the spectrum such as
4.2 -- 6 keV, estimates range from a few percent to half or more
(e.g.~\citealt{Rothschild:2003ev, 2008ApJ...686.1094H}).

The argument over the emission mechanism of the hard tail seems to
have been settled in favor of synchrotron emission.  Electrons in the
10 -- 100 keV energy range required for bremsstrahlung should be
accelerated in the forward or reverse shocks, but Coulomb interactions
ought to cool the electrons efficiently not far behind those shocks,
producing a dip in the spectrum below 100 keV which is not observed \citep{Vink:2008bpa, Vink:2011bka}.
Electron energization by lower hybrid waves in
weak shocks in the remnant's interior might ameliorate this problem \citep{2001ApJ...546.1149L}.
The presumptive site for the production of the very
high-energy electrons required to produce synchrotron emission to 10 keV is
the forward blast wave marked by the faint rim filaments (e.g., \citealt{2004A&A...427..525B}),
though a recent model proposes that almost all emission above 10 keV should come from the reverse shock \citep{Zirakashvili:2014ks}.
Therefore, current theories expect the $>$15 keV emission to be predominantly originate in either the forward shock or the bright
ring.

High spatial resolution observations with XMM and Chandra show that in addition to the clear outer rim and bright ring, there are bright central knots and filaments of soft X-ray emission that are dominated by non-thermal continuum \citep{2004ApJ...613..343D}, with intensities that vary with time \citep{2009ApJ...697..535P}. Whether these features are actually interior to the remnant and evidence for particle acceleration at the reverse shock or knots on the forward shock face seen in projection has remained an open question. It is not known whether the bright ring of the reverse shock, the outer rim of the forward shock, or something else, dominates the emission above 15 keV, which has significant implications for the particle acceleration in the remnant.

\begin{figure}[Hb]
    \includegraphics[width=0.49\textwidth]{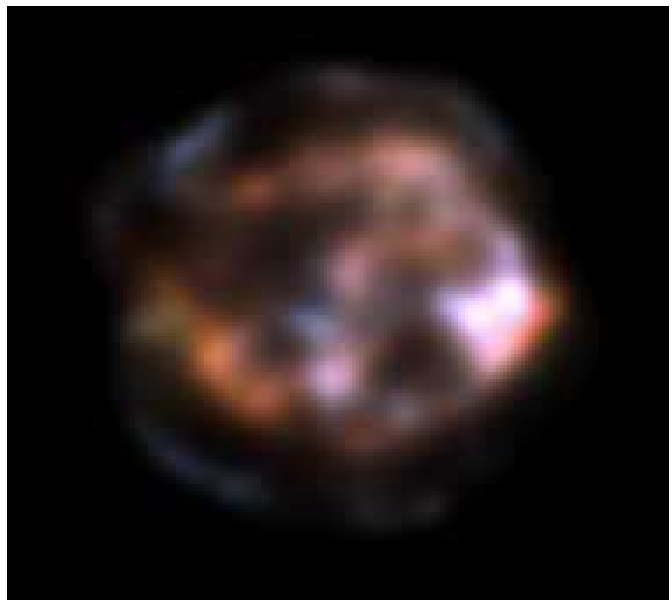}	
  \caption{\footnotesize
  	Deconvolved \NS \ images of Cas A: red (4--6 keV), green (8--10 keV) and blue (10--15 keV). For this and all the images below North is up and East is left. These images are 7.6 arcminutes on a side and all of the color bands have a {\tt sqrt} stretch. As the energy bands increase, the emission is more consistently dominated by the central knots rather than any diffuse emission in the remnant. However, even in the highest energy bands accessible to \NS \ there is still some residual diffuse emission in the center of the remnant not associated with any obvious point sources. See online version for color images.
 }
\label{fig:low_energy}
\end{figure}

In this paper we report on the first spatially resolved hard X-ray images and spectroscopy of Cas A. These collected observations represent over 2.4 Ms of integration time, with the exposure time primarily driven by the study of \TI \ in the remnant \citep{Grefenstette:2014ds}. This results in deep observations that provide unprecedented sensitivity in the hard X-ray band. These observations allow us to make important new measurements: better spectral characterization of the hard continuum (results from previous missions are not entirely consistent with one another), spatial identification of the sites of electron acceleration to the highest
energies, and detailed spectral modeling over most of the energy range of observed X-ray emission. Fully exploiting this new energy band requires extensive analysis combining spatially-resolved information from the radio, soft X-rays, and hard X-rays.
In this paper we present the initial results from the {\it NuSTAR} observations. We compare and contrast our findings with the previously-known characteristics of Cas A and discuss the implications for particle acceleration models. In addition, we provide context for this work by performing a high-level comparison of our findings in the hard X-ray band with previous work in the radio and soft X-ray bands. While the
broadband picture is by no means complete, we expect that this will motivate the future theoretical modeling of the Cas A continuum.

\begin{figure}[Ht]
    \includegraphics[width=0.49\textwidth]{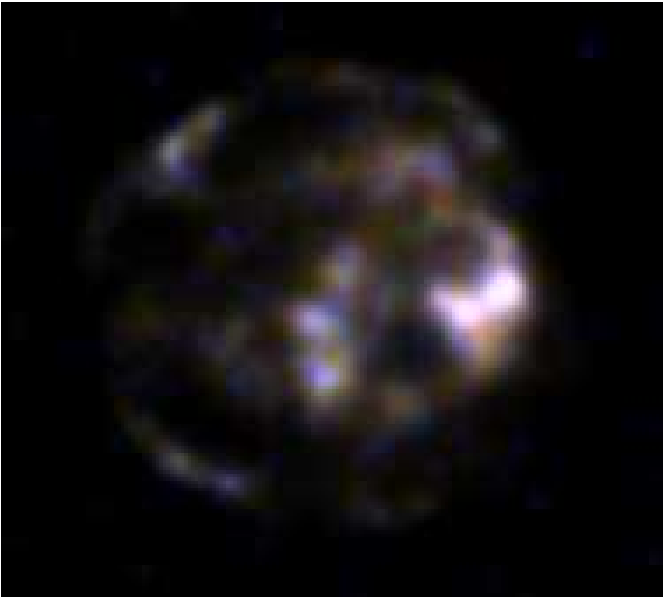}
  \caption{\footnotesize
  	Deconvolved \NS \ images of Cas A: red (15--20 keV), green (20--25 keV), and blue (25--35 keV). See online version for color images.
 }
\label{fig:mid_energy}
\end{figure}

\section{\NS \ Data and Methods}
\label{section:obs}

Cas A was observed by \NS,  a NASA Small Explorer (SMEX) satellite launched on June 13, 2012 \citep{Harrison:2013iq}. \NS \ has two co-aligned X-ray telescopes observing the sky in an energy range from 3--79 keV. The field of view is roughly 12$'$ x 12$'$, with a point-spread function (PSF) with a full-width half maximum of 18$''$ and a half-power diameter of 58$''$.

The observations were completed over the first 18 months of the mission (Table \ref{tab:obs}) with exposures spanning between 120 and 280 ks
in a single pointing. The aim point for the telescope was generally selected to avoid the gaps in the instrument when possible, though later observations were chosen to target specific spatial regions for the study of \TI. Thermal motions of the 10-m mast that separates the \NS \ optics and detectors as well as small variations in the spacecraft pointing introduce natural ``dither" in the pointing pattern, further smoothing the exposure map. There were no roll angle constraints on these observations, so the remnant was observed at different position angles over the course of the observations.

We reduced the \NS \ data with the \NS \ Data Analysis Software (NuSTARDAS) version 1.3.1 and \NS \ CALDB version 20131223. The NuSTARDAS pipeline software and associated CALDB files are fully HEASARC FTOOL compatible and are written and maintained jointly by the ASI Science Data Center (ASDC, Italy) and the \NS \ Science Operations Center (SOC) at Caltech. The NuSTARDAS pipeline generates Good Time Intervals (GTIs) for each observation that exclude periods when the source is occulted by the Earth and when the satellite is transiting the South Atlantic Anomaly (SAA), a region of high particle background. The default detector ``depth cut" is applied to reduce the internal background at high energies. The images, exposure maps, and response files produced by the NuSTARDAS consistently account for the natural thermal motions of the mast.

We produced images in various bandpasses using the {\tt XSELECT} multi-mission FTOOL. Cas A is a bright, extended source so there are no regions in the field-of-view of the telescope that can be used to directly estimate the background in the source region. Instead we model the background and produce energy-dependent background images. We follow the procedure outlined in \cite{Grefenstette:2014ds} and \cite{2014ApJ...792...48W} to estimate the background components and their spatial distributions using the {\tt nuskybgd} IDL suite. In general, the flux from Cas A dominates over the diffuse backgrounds by several orders of magnitude until energies of $\sim$50 keV where the signal becomes comparable to the background.

We combined images from all epochs using {\tt XIMAGE}, taking into account the unique time-dependent exposure and vignetting corrections for each pointing. We chose the energy bands of 4--6 keV, 8--10 keV, 10--15 keV, 15--20, 20--25 keV, 25--35, 35--45, and 45--55 keV for analysis. We omit the 6--8 keV energy band from this work, as it is dominated by Fe line emission from the shocked ejecta and has been explored previously in detail (e.g. \citealt{Hwang:2012gi}). Here we are primarily interested in the continuum emission. For energy bands up to the 25--35 keV we deconvolved the images with the on-axis \NS \ PSF contained in the CALDB and the {\tt max\_likelihood} AstroLib IDL script which is based on a Richardson-Lucy deconvolution for images with Poisson noise. We chose to halt the deconvolution after 50 iterations, since after this the resulting images became sufficiently self-similar. For the 35--45 and 45--55 keV bands there are insufficient statistics to support the image deconvolution, so we instead smoothed the images with top-hat smoothing kernels with radii of 10 pixels ($\sim$25$''$).

\begin{figure*}[ht]
\vbox{
    \includegraphics[width=0.48\textwidth]{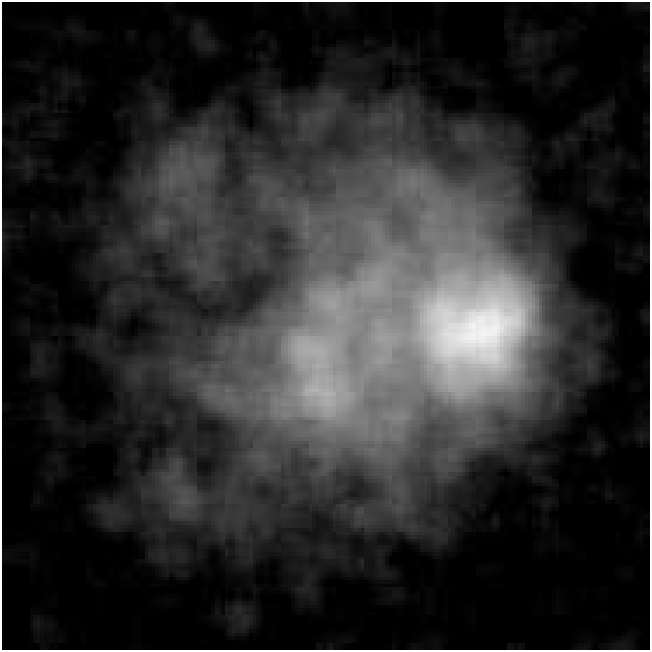}
      \hspace{0.03\textwidth}
    \includegraphics[width=0.48\textwidth]{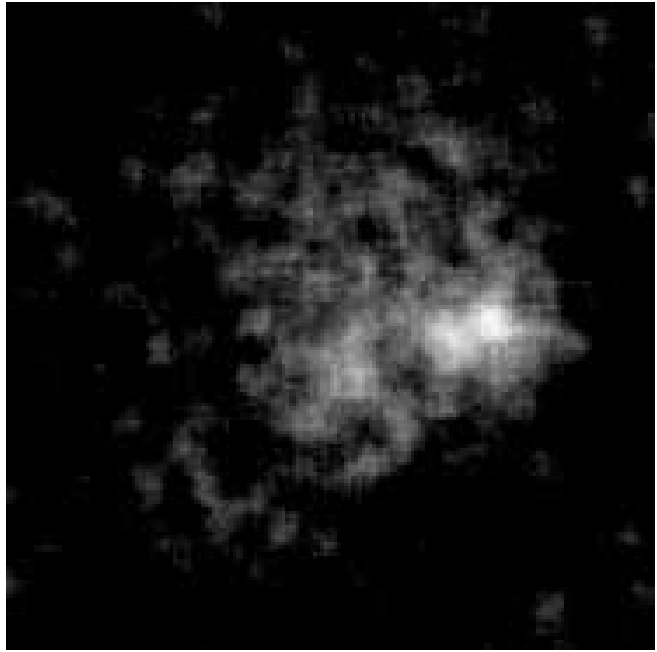}
  
}
  \caption{\footnotesize
  	 \NS \ images of Cas A in the 35--45 keV band (top) and 45--55 keV band (bottom). The images have been smoothed using a top-hat smooth kernels of 10 pixels ($\sim$25$''$) and both have a {\tt sqrt} scaling. The color stretch has been modified for illustrative purposes.
 }
\label{fig:high_energy}
\end{figure*}

To perform spatially resolved spectroscopy we defined standard ds9 region files and used the {\tt nuproducts} FTOOL to extract spectra and produce ancillary response files (ARFs) and redistribution matrix files (RMFs) for each epoch with background spectra simulated using {\tt nuskbygd}. To reduce the complexity of joint-fitting many spectra (14 ObsIds $\times$ two telescopes = 28 spectra), we combined the source pulse-height amplitude (PHA) files, ARFs, and simulated background PHA files using the {\tt addascaspec} FTOOL and combined the RMF using the {\tt addrmf} FTOOL. This results in exposure-weighted RMF and ARF files for each region. Based on many observations of the Crab at different off-axis angles (Madsen et al., in prep), we estimated that the systematic effects caused by combining the spectra in this manner are energy-independent and introduce noise primarily in the normalization of the combined spectrum that is on the order of a few percent. As we are not interested here in in the absolute normalization of the spectra, we ignore this effect below. We also estimate that this introduces a systematic error on the power-law index of $\pm$ 0.01, which is significantly smaller than than the statistical errors quoted below. As we are mostly interested in the non-thermal emission processes, we only fit the data from 15--50 keV. The lower energy bound was chosen to so that the contribution from thermal bremsstrahlung from the shock-heated ejecta studied in detail by {\it Chandra} and {\it XMM-Newton} \ is negligible, while at high energies we chose 50 keV since above this energy the signal from our source regions becomes comparable to the background and mentioned above.

\begin{figure}[ht]
\begin{center}
    \includegraphics[width=0.5\textwidth]{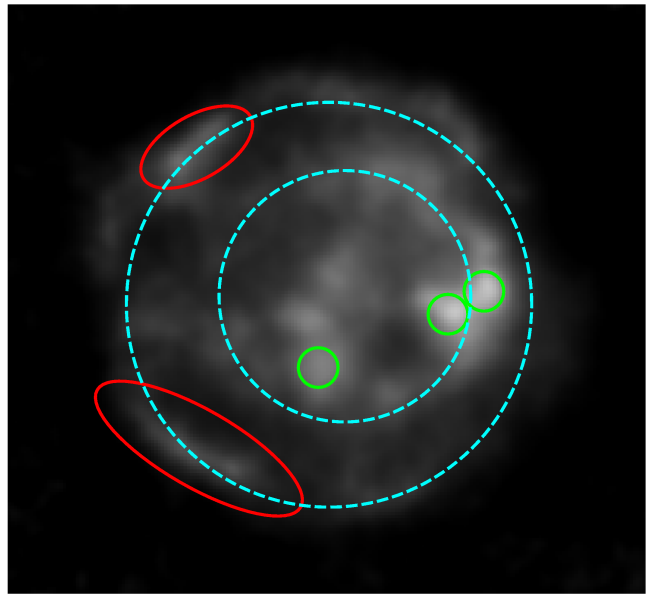}	
  \caption{\footnotesize
  	Spectral extraction regions. The 15--20 keV \NS \ image of Cas A shows an 8$\times$8 arcminute region around the remnant shown with an aggressive stretch to highlight the residual diffuse emission through the remnant. Also shown are the extraction regions used in the spectral analysis: the green circles are the representative central knot regions (all of which have similar spectra), while the red ellipses are the northeast and southeast exterior filaments. These latter regions roughly correspond to exterior filaments of \cite{2009ApJ...697..535P} and filaments 1, 2, 3, and 9 of \cite{2010ApJ...714..396A}. The cyan dashed circles are the approximate locations of the forward and reverse shocks from \cite{Gotthelf:2001bn}. See online version for color images.}
\label{fig:regions}
\end{center}
\end{figure}

%

\begin{table}
  \caption{\NS \ Observations}
  \label{tab:obs}
  \begin{center}
    \leavevmode
    \begin{tabular}{lrc} \hline \hline              
  OBSID     	&	 Exposure   & 	UT Start Date    \\ \hline 
40001019002 &  294 ks & 2012 Aug 18 \\
40021001002 & 190 ks & 2012 Aug 27 \\
40021001004 & 29 ks &  2012 Oct 07 \\
40021001005 & 228 ks & 2012 Oct 07 \\
40021002002 & 288 ks & 2012 Nov 27 \\
40021002006 & 160 ks & 2012 Mar 02 \\
 40021002008 & 226 ks & 2012 Mar 05 \\
40021002010 & 16 ks & 2012 Mar 09 \\
40021003003 & 13 ks & 2013 May 28 \\
40021003003 & 216 ks & 2013 May 28 \\
40021011002 & 246 ks & 2013 Oct 30 \\
40021012002 & 239 ks & 2013 Nov 27 \\
40021015002 & 86 ks & 2013 Dec 21 \\
40021015003 & 160 ks & 2013 Dec 23 \\
  \hline
  Total & $\approx$ 2.4 Ms & \\ \hline
    \end{tabular}
  \end{center}
\end{table}

\section{Results}

\subsection{NuSTAR Imaging}

While the spatially integrated hard X-ray emission from Cas A has been measured by a number of instruments, \NS \ provides the first opportunity to spatially resolve the emission above 15 keV (Figures \ref{fig:low_energy} -- \ref{fig:high_energy}). We adopt the terminology of ``exterior" emission features to be those seen near the outside of the remnant when seen in projection on the sky and ``central" emission features to be those seen toward the center of the remnant when seen in projection on the sky. We also adopt a description of ``knots" of emission as those regions that are unresolved by \NS \ and ``filaments" as those that appear to be linearly extended regions of emission.

We find that above 15 keV the morphology of the remnant begins to deviate from the emission below 12 keV observed by \CHAN, \XMM, and \SUZ. While the outer filaments (i.e. the  ``thin rim"  of the forward shock) are clearly visible in the \NS \ images, the emission is dominated by two central unresolved knots in the west. These western central knots dominate the hard X-ray emission above 15 keV, which is broadly consistent with the results from \cite{2008ApJ...686.1094H}, who found a global east/west asymmetry in the hardness ratio of the remnant based on data from \CHAN, implying that the west should be brighter at higher energies. However, we note that though the central western knots dominate the emission, the tricolor image above 15 keV (Figure \ref{fig:mid_energy}) demonstrates that the exterior filaments are harder (bluer) than the central knots, a fact we explore quantitatively below.

\begin{figure}[ht]
\begin{center}
    \includegraphics[width=0.49\textwidth]{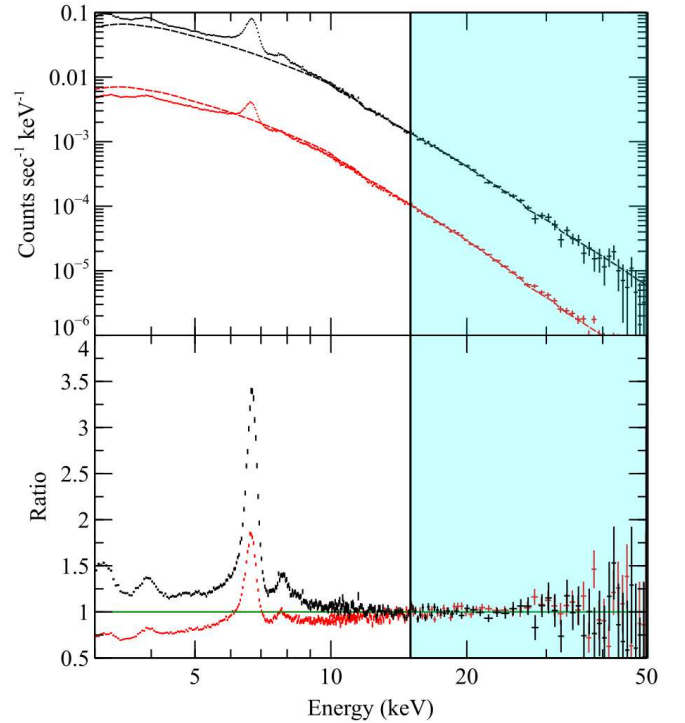}	
  \caption{\footnotesize
  	Combined spectra of exterior filaments and central knots of Cas A. The combined spectrum from the northeast and southeast external filaments is shown in black (upper curve), while the combined spectrum of the central knots is shown in red (lower curve). The spectrum of the knots has been artificially offset downwards by a factor of 10 for clarity in the top frame. The shaded region to the right is the 15--50 keV band used to fit the power-law component, which has then been extrapolated down to 3 keV. Across the 15--50 keV band the spectrum is well fit by a single power-law. The normalization of the power-law is allowed to vary between the different regions, though the index is tied together for the two exterior filament regions and the three central knot regions. Best-fit parameters are given in Table \ref{table:fits}. The central knots have a softer power-law spectrum than the exterior filaments. When extrapolated to low energies, the central knots overpredict the observed emission while the exterior filaments underpredict the observed emission as demonstrated in the lower panel, which shows the ratio of the data to the models. See online version for color images.
	\vspace{-15pt}
}
\label{fig:spectra}
\end{center}
\end{figure}

\subsection{Spatially Resolved Spectroscopy}
\label{sec:spatial-spec}

With \NS \ we can separately analyze the non-thermal continuum originating from different spatial regions of the remnant. Figure \ref{fig:regions} shows the 8--10 keV \NS \ image along with the extraction regions that we used for this work.

We find that all of the regions are well fit by unbroken power-law models across the 15--50 keV band. The exterior filaments (the regions labeled ``northeast" and ``southeast" in Figure \ref{fig:regions}) have similar spectra, with power-law indices of $\Gamma \sim -3.06$, while the interior bright knots (the ``knot" regions in Figure \ref{fig:regions}) also have similar spectra, but in general have a softer power-law index of $\Gamma \sim -3.35$. We fit all of the regions independently but, for clarity, show the combined spectra and best-fit models for the exterior and interior regions in Figure \ref{fig:spectra} while the best-fit model parameters and 2-$\sigma$ error ranges are given in Table \ref{table:fits}. 

At soft X-ray energies ($<$10 keV), the presence of lines associated with ionized S, Fe, and Ni (as well as lighter species below the calibrated \NS \ band) indicate the presence of a hot thermal plasma. We do not attempt to model these thermal plasmas since we know from analysis with \CHAN \ that the plasma properties may vary on $\sim$arcsecond spatial scales \citep{Stage:2006eza, Hwang:2012gi} and the relatively large \NS \ regions will sample many regions with different physical parameters. However, we note that when we extrapolate the power-law fits from the 15--50 keV band to soft X-ray energies ($<$10 keV) we see different behavior for the exterior filaments and the central knots; the power-law model for the exterior filaments under-predicts the observed spectrum at soft X-ray energies while the power-law model for the central knots over-predicts the observed spectrum. We discuss the implications of this below.


\begin{table}[ht]
  \caption{\NS \ power-law fits over the 15--50 keV band}
  \label{tab:threeband_fits}
  \begin{center}
    \leavevmode
    \begin{tabular}{lc} \hline \hline              
 	 Region     	&  Photon Index (2-$\sigma$ Error Range)   \\ \hline 
	 Central Knots & $-3.35$ (3.29--3.41) \\
	 Outer Filaments & $-3.06$ (2.98--3.13) \\
    \end{tabular}
    \vspace{30pt}
  \end{center}
  \label{table:fits}
\end{table}

\begin{figure*}[ht]
\begin{center}
    \includegraphics[width=0.31\textwidth]{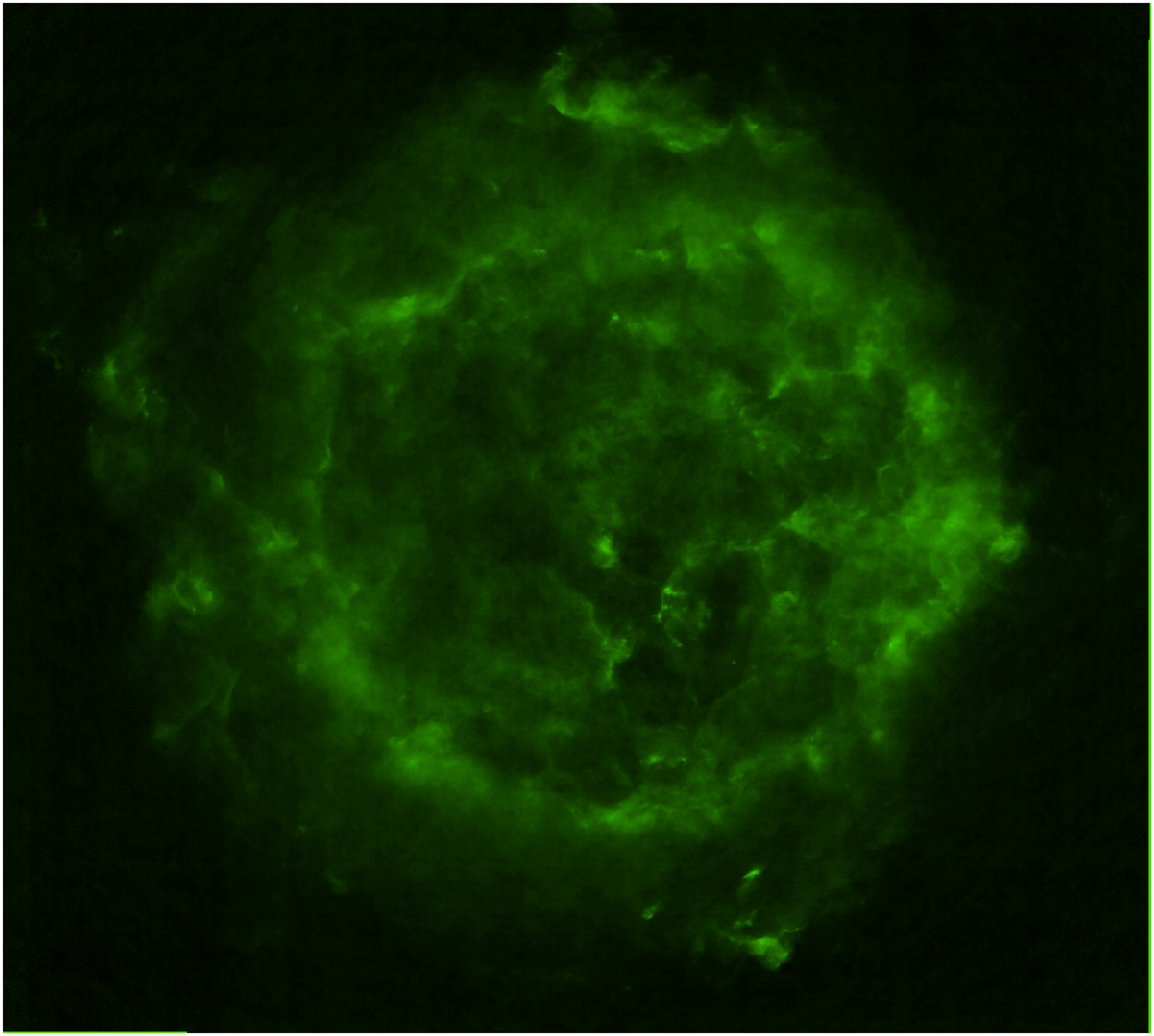}	
    \hspace{0.005\textwidth}
    \includegraphics[width=0.31\textwidth]{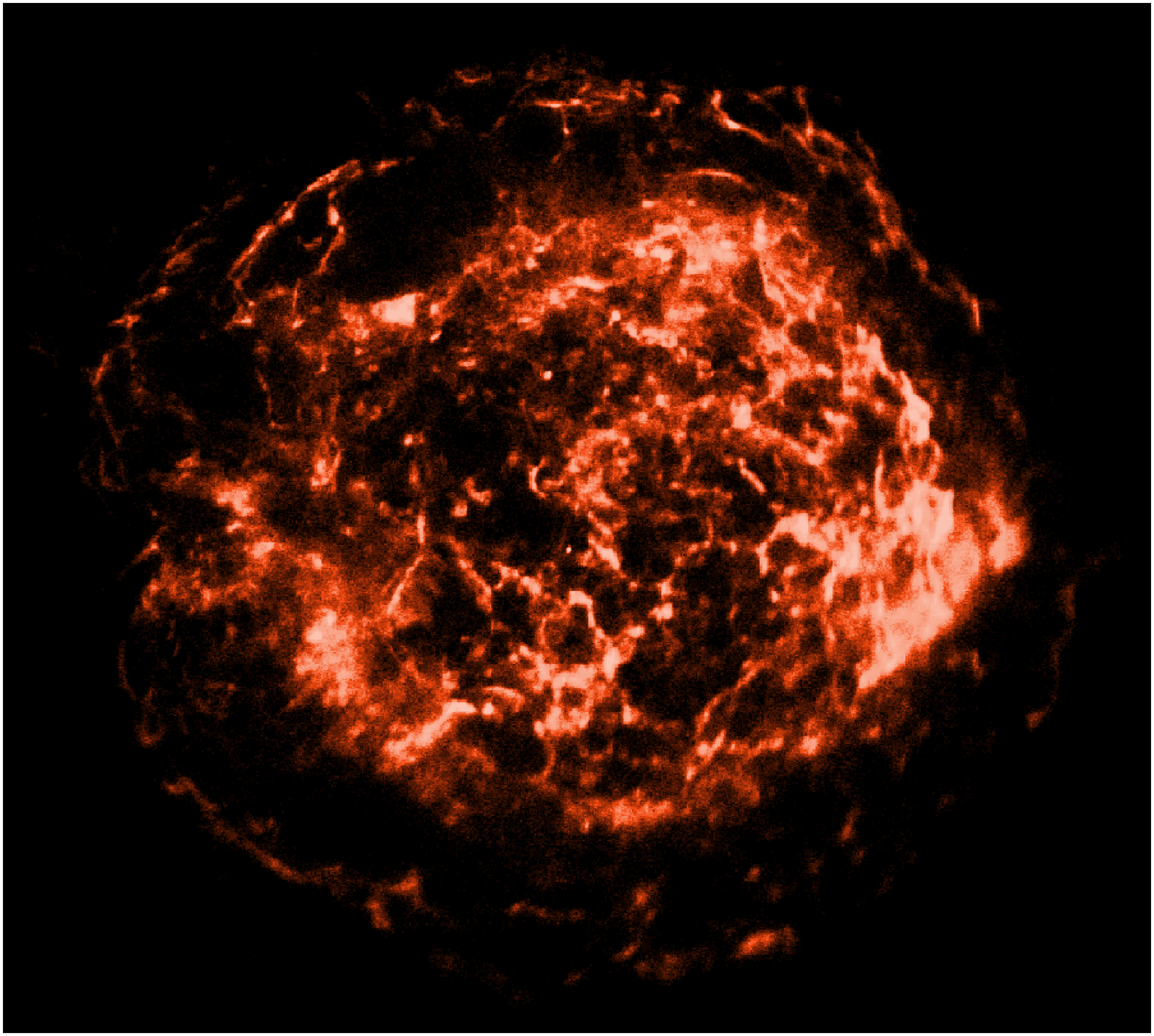}
    \hspace{0.005\textwidth}   
    \includegraphics[width=0.31\textwidth]{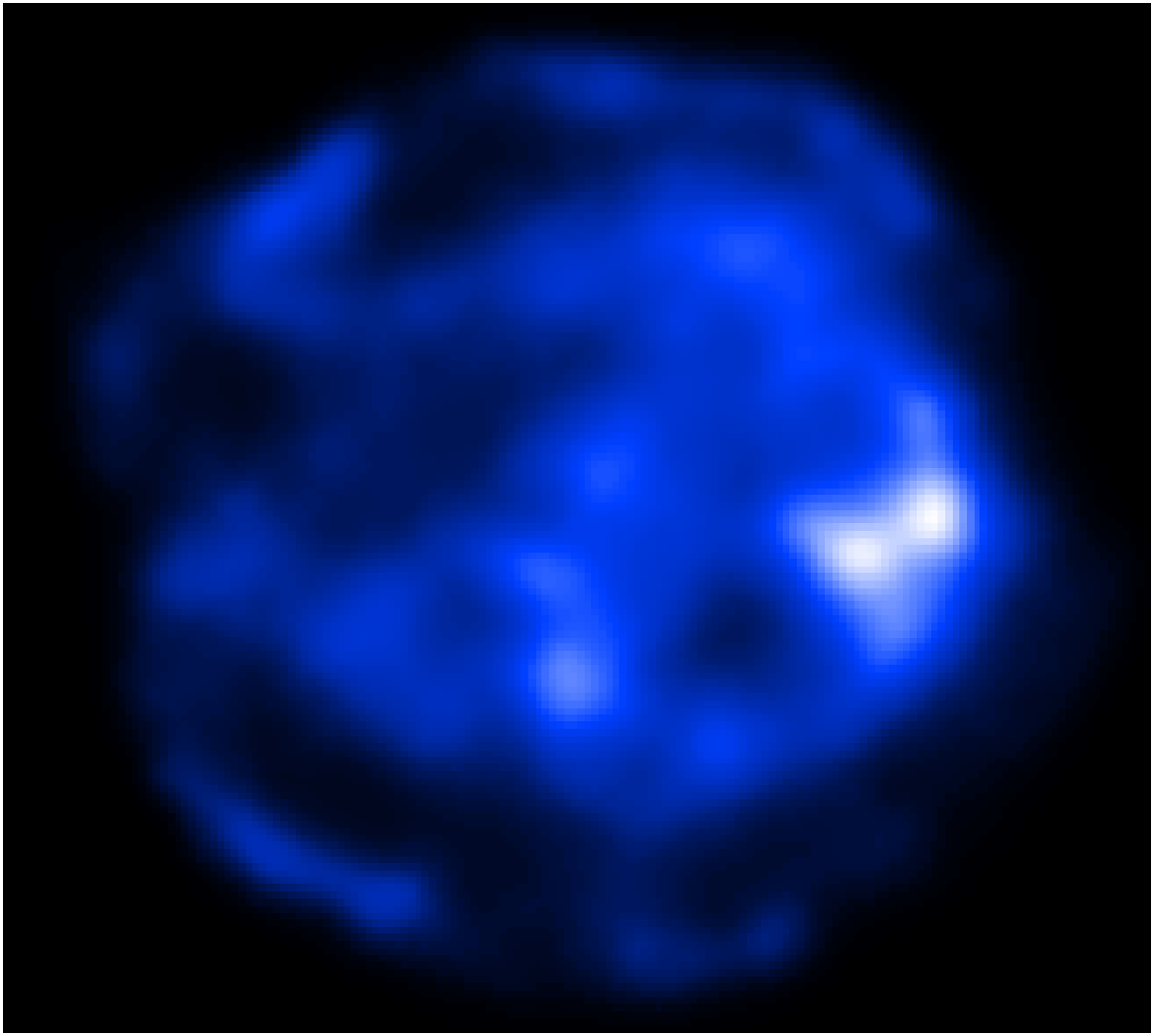}
  \caption{\footnotesize
  	\NS \ continuum image of Cas A. The 10--15 keV \NS \ shows the image after it has been deconvolved with the PSF. The image has been been aggressively stretched to highlight the dimmer, diffuse emission from the remnant. See online version for color images.}
\label{fig:rachnu_soft_tripanel}
\end{center}
\end{figure*}

\subsection{Multiwavelength Comparisons}

Figure \ref{fig:rachnu_soft_tripanel} shows a comparison between the radio (6 cm intensity maps obtained with the VLA), soft X-rays (4--6 keV continuum images taken with \CHAN \ from \cite{Hwang:2004dua}), and the 10--15 keV hard X-ray observed by \NS.

\subsubsection{Hard X-rays and Radio}

\begin{figure*}[ht]
    \includegraphics[width=0.49\textwidth]{figure6a.eps}	
    \hspace{0.005\textwidth}
    \includegraphics[width=0.49\textwidth]{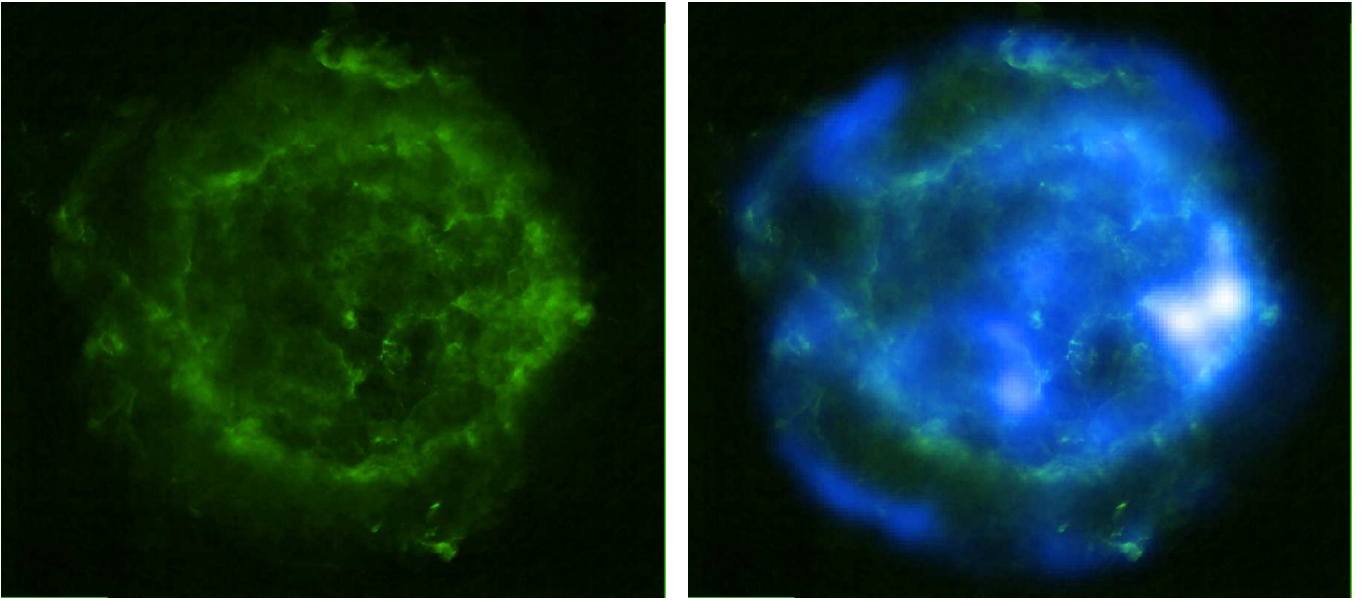}
  \caption{\footnotesize
  	Comparison of the VLA 6 cm radio and \NS \ continuum images of Cas A. Left: VLA 6 cm image. Right: The overlay of the VLA image with \NS \ image from Fig \ref{fig:rachnu_soft_tripanel}. See online version for color images.}
	\label{fig:ranu_soft}
\end{figure*}

The hard X-ray and radio morphologies of Cas A are substantially different (Figure \ref{fig:ranu_soft}). The outer filaments in the northeast and southeast visible in NuSTAR are coincident with the edge of faint radio emission in the VLA images, but the radio morphology is not of thin tangential filaments but simply is simply described by a broad plateau. Any filamentary structure could be unobserved due to a lack of dynamic range in the radio images (i.e. the radio emission from this region is faint compared to the emission in the bright ring) or to a change in the shape of the continuum extending to the radio. We do not see any enhancements in the hard X-ray images near the center of the remnant where we see the bright ring in the radio emission. Instead, we see that in the eastern half of the remnant there is little hard X-ray emission associated with the bright ring, while in the western half of the remnant the bright central knots appear to be located near the bright ring.

\subsubsection{Hard X-rays and Soft X-rays}

\begin{figure*}[ht]
    \includegraphics[width=0.49\textwidth]{figure6b.eps}	
    \hspace{0.005\textwidth}
    \includegraphics[width=0.49\textwidth]{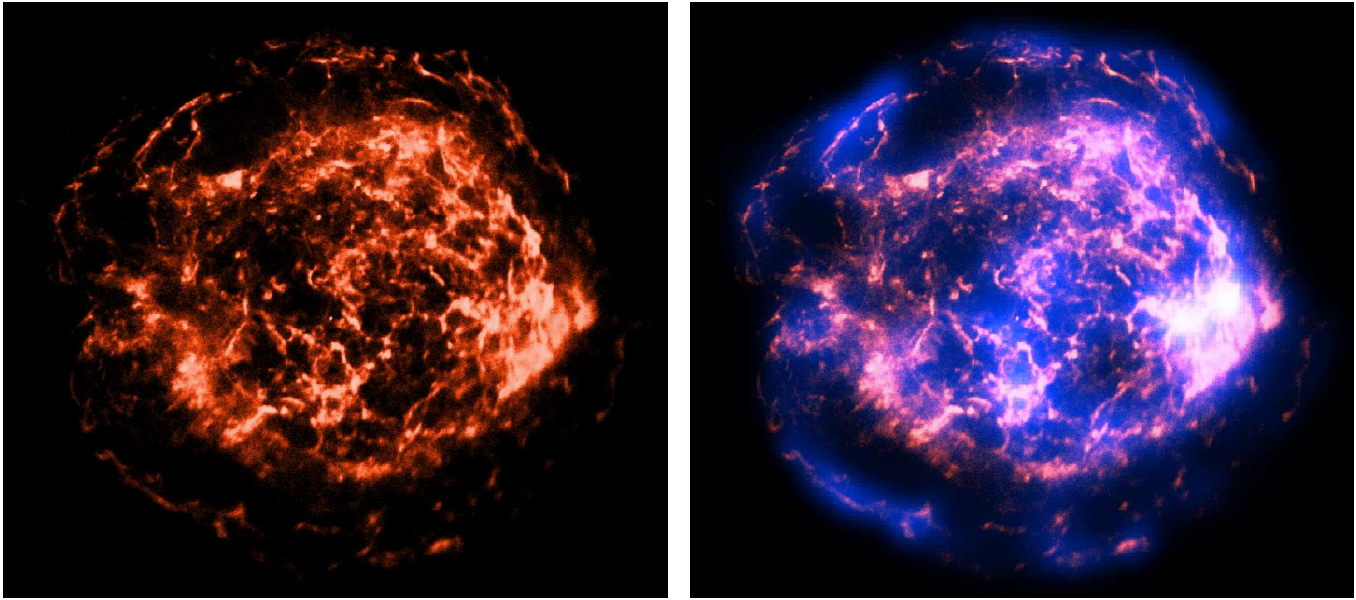}
  \caption{\footnotesize
  	Comparison of the \CHAN \ and \NS \ continuum images of Cas A. Left: \CHAN \ 4--6 keV image. Right: the overlay of the \CHAN \ image with \NS \ image from Fig \ref{fig:rachnu_soft_tripanel}. The images have been stretched to highlight the dimmer, diffuse emission from both satellites. See online version for color images.}
\label{fig:chunu_soft}
\end{figure*}

We find that the \CHAN \ and \NS \ images (Figure \ref{fig:chunu_soft}) generally agree on large spatial scales up to $\sim$10--15 keV, but show differences in the higher energy bands observed by \NS. Up to 15 keV, the exterior filaments in the forward shock of Cas A in the north, northeast, south, and southeast seen with \CHAN \ are all clearly visible in the $<$15 keV~\NS~images, as are some of the interior emission which is likely to be residual flux from hot thermal plasmas leaking into the \NS \  band. It is in the relative intensity of the central knots that we note the major differences, with the western knots dominating much of the hard X-ray flux in the 10--15 keV band in \NS \ while they appear relatively unremarkable in the \CHAN \ images, even when steps are taken to reduce the impact of the bright thermal emission in Cas A (e.g. using imaging tomography as per \cite{2004ApJ...613..343D}). This difference becomes more apparent in the images at higher energies (Figures \ref{fig:mid_energy} and Figure \ref{fig:high_energy}), with the central knots in the west dominating the emission above 15 keV and the bright ring completely disappearing in the these energy bands. At the highest energies (Figure \ref{fig:high_energy}) the emission appears to be mostly attributable to several of the bright knots, with the outer rims fading away. Several of the central knots are also near regions noted by \cite{2009ApJ...697..535P} and \cite{2008ApJ...677L.105U} to show variability in the 4--6 keV continuum emission on timescales of a year, which we discuss below.

\subsubsection{Radio, Soft, and Hard X-rays}

\begin{figure}[hb]
\begin{center}
    \includegraphics[width=0.49\textwidth]{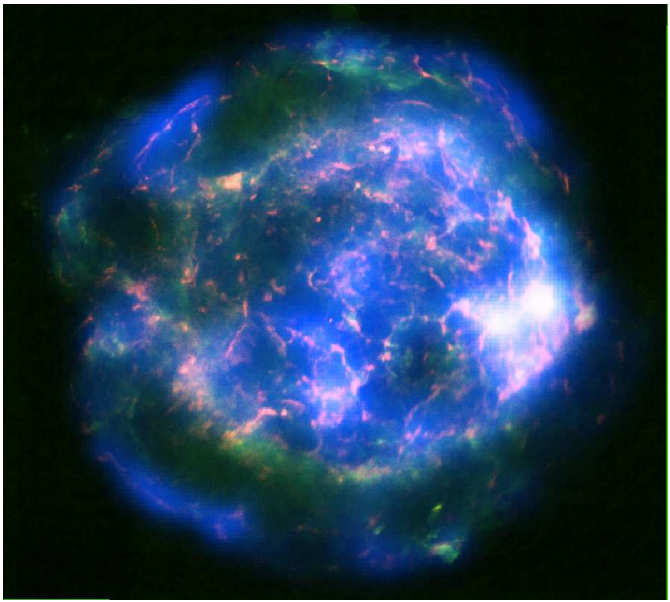}	
  \caption{\footnotesize
  	Comparison of the \CHAN, VLA, and \NS \ continuum images of Cas A. Red: \CHAN \ 4--6 keV data from the left panel of Figure \ref{fig:chunu_soft}; Green: VLA 6 cm data from the left panel of Figure \ref{fig:ranu_soft}; Blue: \NS \ 10--15 keV data from Figure \ref{fig:rachnu_soft_tripanel}. See online version for color images.}
	\label{fig:rachnu_soft}
\end{center}
\end{figure}

\begin{figure*}[ht]
\begin{center}
\hbox{
    \includegraphics[width=0.23\textwidth]{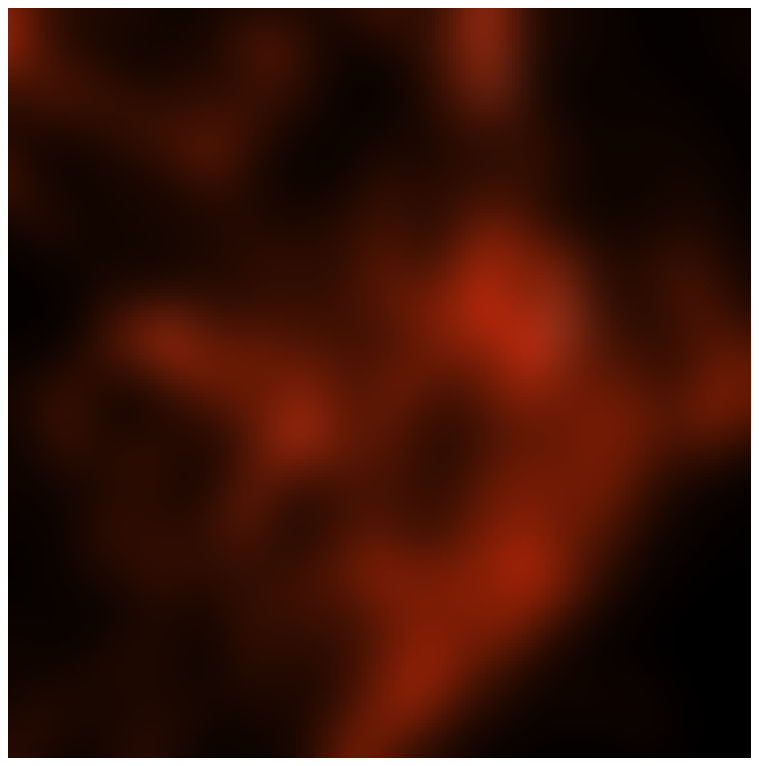}	
    \hspace{0.005\textwidth}
    \includegraphics[width=0.23\textwidth]{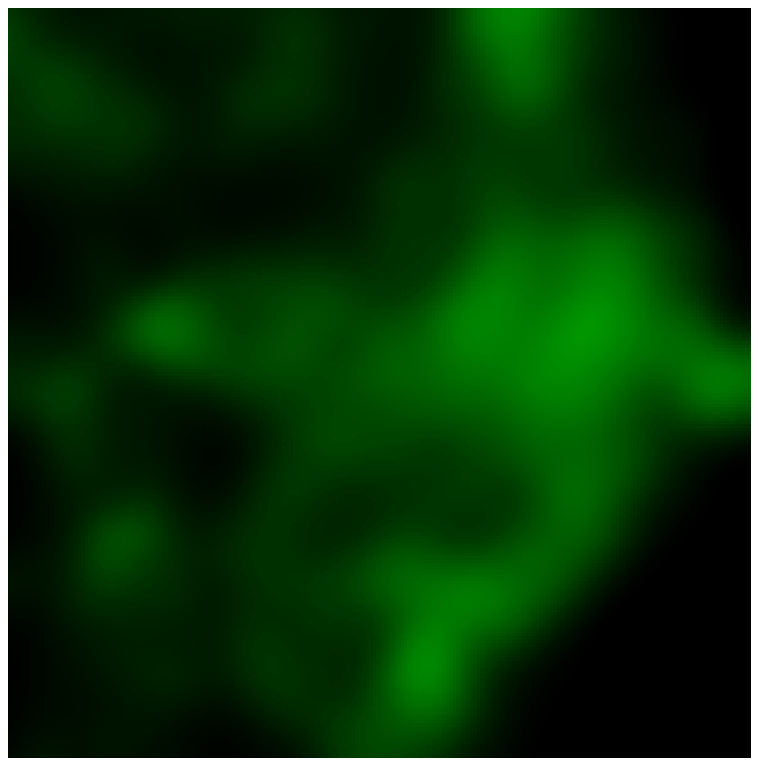}
    \hspace{0.005\textwidth}   
    \includegraphics[width=0.23\textwidth]{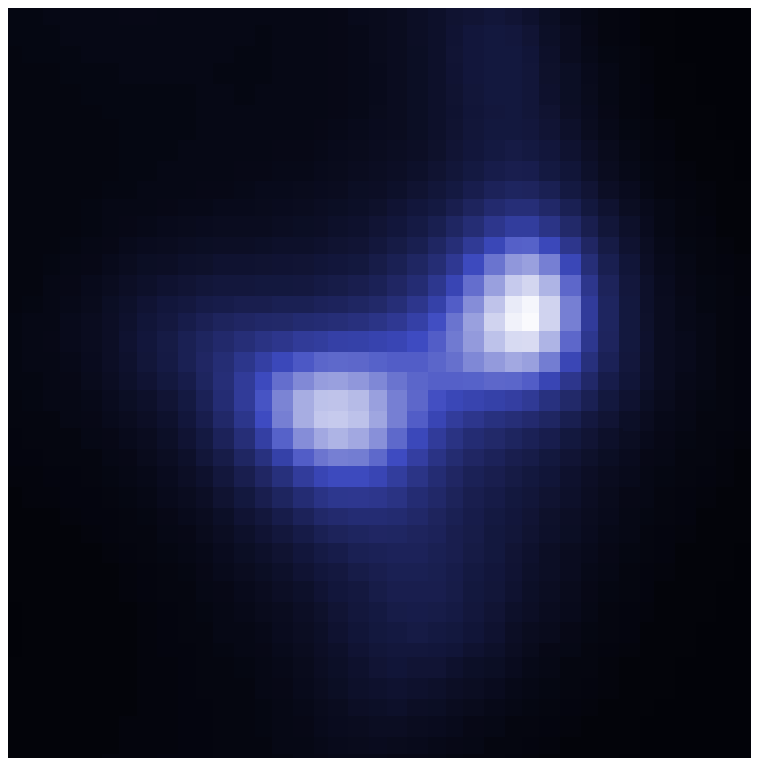}
    \hspace{0.005\textwidth}   
    \includegraphics[width=0.23\textwidth]{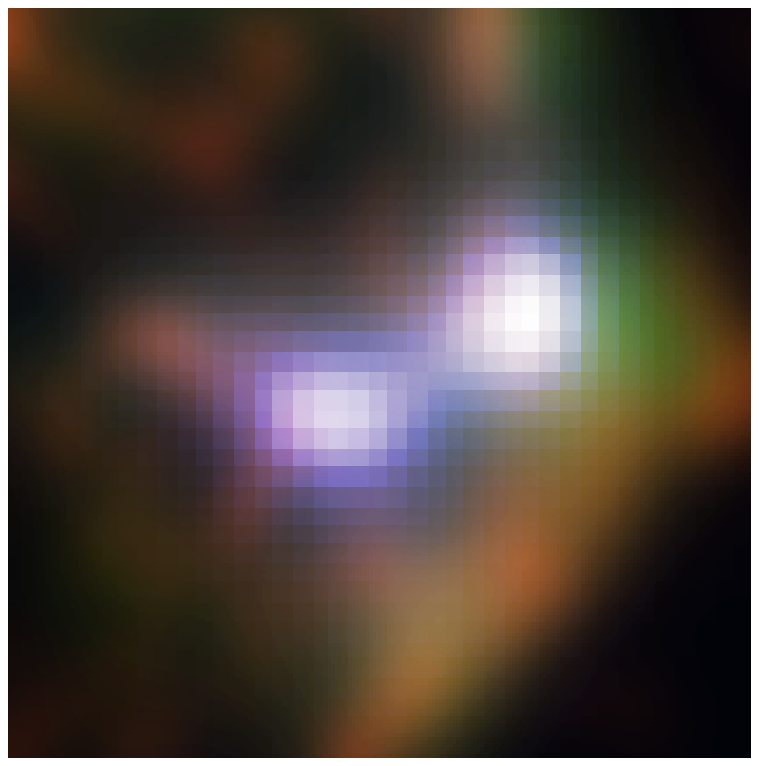}

}
  \caption{\footnotesize
  	Comparison of the \CHAN, VLA, and \NS \ images of Cas A, zoomed in on the bright western knots. The color scheme is the same as for Fig \ref{fig:rachnu_soft}. Here the \CHAN \ and VLA data have been smoothed to approximately the same resolution as the \NS \ 10--15 keV data and all of the images have been stretched for ease of comparison. See online version for color images.}
\label{fig:smallhard}
\end{center}
\end{figure*}

Figure \ref{fig:rachnu_soft} shows the overlay of the radio, soft, and hard X-ray images. In the particular case of the western knots, we find that the two hard X-ray knots observed by \NS \ differ from both the soft X-ray and radio images. Figure \ref{fig:smallhard} shows a zoomed-in view around the two western knots, with the \CHAN \ and the VLA observations smoothed to approximately the same spatial scales as the \NS \ deconvolved images. We see that the radio is clearly strongest near the western of the two knots, \CHAN \ shows some flux near the eastern knot but is still dominated by the western knot, while \NS \ observes two knots of nearly equivalent brightness. These central knots are spatially consistent with two bright knots in high-resolution maps of the radio luminosity and radio spectral index (\citealt{Anderson:1996bk} and \citealt{DeLaney:2014kj}), with the brighter of the two knots associated with a region of steeper radio spectral index while the other is associated with a region of flatter radio spectral index. The fact that \NS \ sees comparably bright X-ray emission in both of these regions is puzzling.

\subsubsection{keV, GeV, and TeV emission}

\begin{figure}[ht]
\begin{center}
    \includegraphics[width=0.5\textwidth]{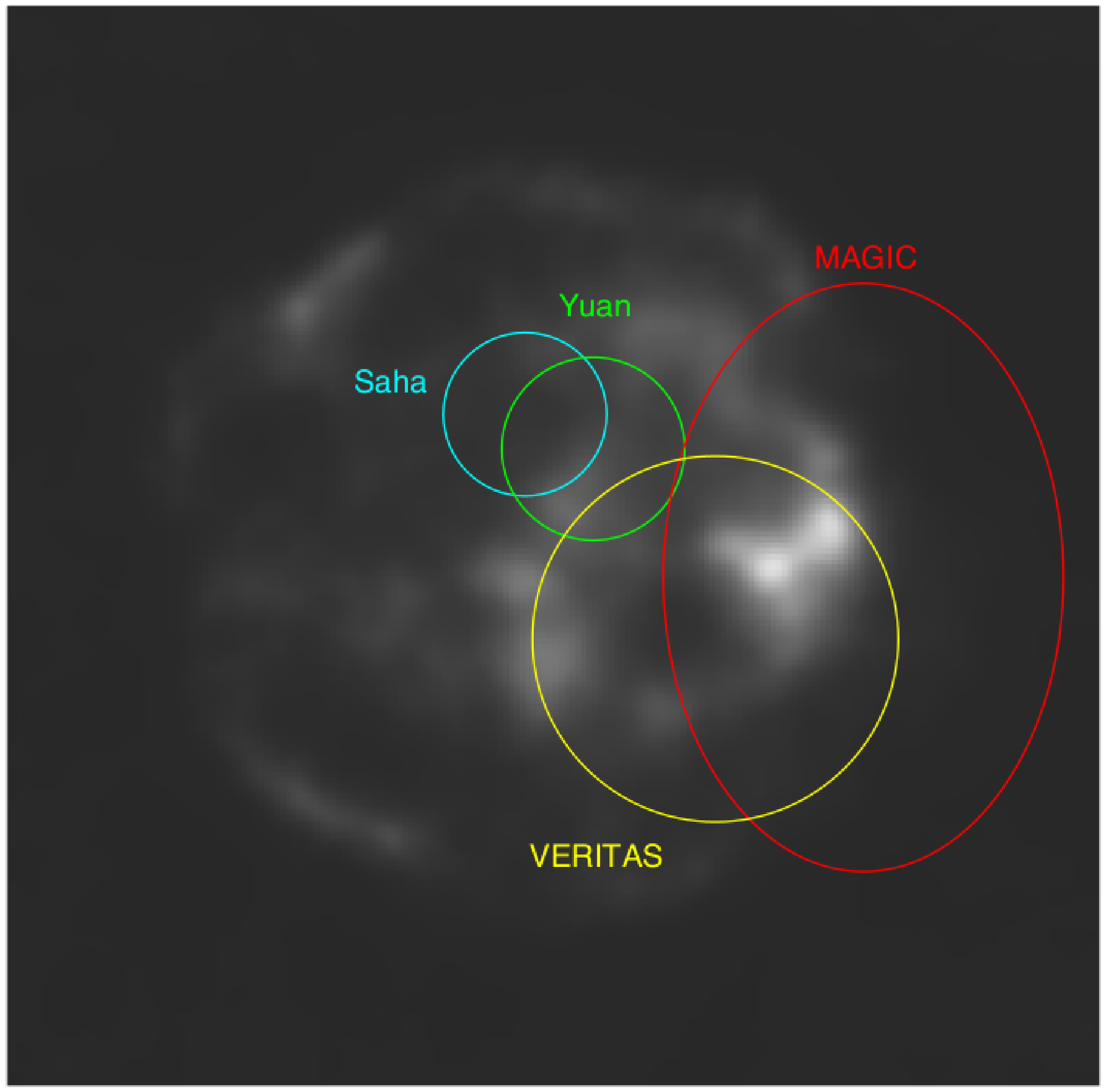}	
  \caption{\footnotesize
  	Comparison of the \NS \ 15--20 keV 8x8 arcminute image with the centroid locations from two Fermi reductions (green circle: \citealt{2013ApJ...779..117Y}; cyan circle: \citealt{Saha:2014ep}), VERITAS (yellow circle: \citealt{2010ApJ...714..163A}), and MAGIC (red ellipse: \citealt{2007A&A...474..937A}). Note that this image is rotated with respect to the comparison of the VHE centroids to the radio image shown by \citealt{2013ApJ...779..117Y} as those authors present the data in Galactic coordinates. All the GeV and TeV regions have a radius that corresponds to the $1\sigma$ statistical errors added in quadrature with the systematic errors. See text for details. See online version of color images. 
}
\label{fig:tev}
\end{center}
\end{figure}

There has been improved work over the last few years localizing and studying the GeV to TeV emission from Cas A. This is important for understanding the acceleration of cosmic rays up to the ``knee" of the cosmic ray spectrum. MAGIC \citep{2007A&A...474..937A} and VERITAS \citep{2010ApJ...714..163A} have observed TeV emission that peaks in the western half of the remnant, while \cite{2013ApJ...779..117Y} show that the {\it Fermi} GeV emission peaks more towards the center of the remnant. We compare and contrast the centroids of the TeV and GeV emission with features in the 15--20 keV band (Figure \ref{fig:tev}). At the 90\% level the GeV and TeV centroids are consistent with each other and could be associated with any part of the remnant. However, if we speculatively use the stricter 1-$\sigma$ error limits then we note that the GeV centroid is located closer to regions of the remnant known to be bright in the infrared while the TeV centroid is closer to a bright region in the NuSTAR band.

We reproduce the data represented in Figure \ref{fig:tev}, as they are often reported in different coordinate systems and are shown with different confidence contours. We adopt decimal degrees in $(\alpha, \delta)$ as our coordinate system and show $1\sigma$ statistical errors added in quadrature with the systematic errors quoted in the literature. \citet{2007A&A...474..937A} find the TeV emission centered at (350.79, 58.81), with asymmetric errors in the $\alpha$ direction of ($0.003_{stat} + 0.001_{sys})$ hours, which, when added in quadrature and projected into the tangent plane at a declination of 58.81 degrees, corresponds to an error estimate of 0.0245 degrees. The $\delta$ error is ($0.03_{stat} + 0.01_{sys})$ degrees, or 0.036 degrees, when added in quadrature. For VERITAS, the centroid shown in \citet{2010ApJ...714..163A} is (350.825, 58.8025) with symmetric errors in the tangent plane of $0.01_{stat} + 0.02_{sys}$ degrees, or 0.022 degrees when added added in quadrature. The {\it Fermi} centroids reported in the literature vary slightly in the literature owing (we assume) to a difference in the data volume used by two authors (\citealt{2013ApJ...779..117Y} and \citealt{Saha:2014ep}). The two reductions appear to largely be consistent with one another: \citet{2013ApJ...779..117Y} gives the location in Galactic coordinates of (111.74, -2.12) corresponding to $(\alpha, \delta)$ of (350.2876, 58.5513), while \citet{Saha:2014ep} shifts the centroid slightly to the northwest at coordinates (350.87, 58.83). In both cases the errors are symmetric $(0.01_{stat} + 0.005_{sys})$, or 0.0112 degrees.

\section{Discussion}

\subsection{Hard X-ray Morphology}

Previous observations of the non-thermal continuum emission in the \CHAN \ and \XMM \ X-ray bands suggest a non-thermal thin rim associated with the ``forward shock" \citep{Gotthelf:2001bn} with some additional filamentary structures through the center of the remnant \citep{2000ApJ...528L.109H}. Based on \XMM \ data, \cite{2001A&A...365L.225B} conclude that the non-thermal emission traces the softer thermal component and that the emission does not originate from a few localized regions. We have shown for the first time that the $>$15 keV flux from Cas A is dominated by neither the forward shock nor the bright ring (presumably the reverse shock). Instead, we find that, contrary to all expectations, the $>$15 keV emission is dominated by the emission from several bright knots near the center of the remnant.

The nature of the central emission in Cas A has been a long-standing debate (e.g., see  \citealt{2008ApJ...677L.105U} and \citealt{2008ApJ...686.1094H}). If the central knots of emission were actually simply the exterior filaments from the outer rim of the remnant projected onto the center then we would expect the spectral shape of the central knots to be similar to the spectral shape of the exterior filaments. Our observations have conclusively shown that the spectra from the two regions are quantitatively different. This points to some systematic shift between the exterior and central spatial regions, be it a difference in the underlying electron population, a difference in how the electrons are being accelerated, a difference in how the electrons are producing X-rays, or all of these.

If the central (western) knots are located at the forward shock front then the difference in the power-law index between the different spatial regions must be attributable to some change in the physical environment (e.g. higher magnetic fields, different compression ratios in the shock front, etc). While we cannot explicitly rule out this possibility, we find it suggestive that the individual outer filaments share a similar spectrum while the individual central knots share a (different) similar spectrum. If all of the emission were coming from the forward blast wave then there would need to be some large scale physical differences between the outer rim emission seen in the northeast and southeast and the central emission seen in the west. Morphologically the central knots also appear point like, which is not what we would expect if we were seeing a face-on version of the filamentary structures on the (eastern) outer edge of the remnant. Finally, it seems unlikely that a viewing-angle dependence for the emission mechanism would conspire to alter the observed spectral index for the outer rims as compared to the central knots. We thus conclude that we are observing two relatively independent electron populations and therefore argue that the central knots are, in fact, located in the interior of the remnant.

This leaves the unsolved problem of the origin of the electrons responsible for this emission in the interior of the remnant. \cite{2008ApJ...686.1094H} argue, with some assumptions about the turbulence of the magnetic field, that electrons accelerated at the forward shock could not have advected to the interior of the remnant over timescales consistent with the age of the remnant and before the electrons would cool via synchrotron losses. A related observational fact is that both the emission near the exterior filaments and internal knots have been shown to vary on $\sim$yearly timescales \citep{2007AJ....133..147P,2008ApJ...677L.105U}. It is not yet clear whether these variations are caused by changes in the environment surrounding a (relatively static) population of electrons or whether some unknown mechanism is accelerating electrons in the interior of the remnant. If the latter were true, then we may also need to revisit non-thermal bremsstrahlung as a possible emission mechanism at least for the interior emission. \cite{Zirakashvili:2014ks} has put forward the idea that the secondary electrons from the radioactive decay of \TI \ may be important for seeding the relativistic electron population responsible for the non-thermal X-ray emission, but as the morphology observed by \NS \ is dramatically different than the morphology of the \TI \ in Cas A \citep{Grefenstette:2014ds} we do not think that this scenario is likely. We also note that if the electrons are being accelerated in the interior of the remnant at the reverse shock, we do not understand why they are only being accelerated at particular locations near the reverse shock. Again, perhaps this is driven by variation in the nature of the ejecta or the gas in the interior of the remnant near the reverse shock. As of now, we do not see a definitive solution for this problem and leave the origin of this emission as an open mystery. \\

\subsection{The nature of the non-thermal emission}

The morphology of the $>$15 keV emission is puzzling and does not appear to lend itself to a simple interpretation. Models which only produce accelerated electrons near a uniform forward shock (e.g. that only predict non-thermal emission in a thin rim surrounding the supernova remnant) are disfavored without any of the complications due to the presence of the thermal flux in the 4--6 keV band from the shocked ejecta. The non-thermal emission (both in the radio as well as in the soft and hard X-ray bands) implies that the electrons producing the $>$15 keV emission should be the high energy tail of the same electrons producing the radio emission. Therefore the hard X-ray morphology should follow some of the large-scale features of the radio map. We find that this not to be the case and do not have a plausible explanation for why the non-thermal X-ray and radio maps differ significantly. A detailed, spatially resolved comparison of the radio spectral index and flux with the hard X-rays is required, but is beyond the scope of our present work.


It is still possible that the emission process for the bright knots is
not synchrotron emission from 10 -- 100 TeV electrons at all, but
rather bremsstrahlung from electrons with only somewhat suprathermal
energies of 10 -- 100 keV. \cite{2001ApJ...546.1149L} suggested that
lower hybrid waves at low Mach-number reflected shocks in the interior
could produce such electrons. While the primary objection to such models (e.g.
\citealt{Vink:2008bpa}), that the electrons above thermal energies would
rapidly cool due to Coulomb losses, would still need to be met, the localization
of the emission into two bright knots may indicate some unique conditions that hold
in the knots but not in the rest of the remnant.
The predicted spectral shape of such emission can be
adjusted somewhat by varying the electron Alfv\'en velocity, but the
particular calculations of \cite{2001ApJ...546.1149L} all give a cutoff above
$\sim$100 keV. If the continuum reported by {\it Beppo-SAX} out to 300 keV
were confirmed then this could disfavor this picture.

The particular case of the bright western knots may be a connection between the X-rays and the radio bands. \cite{2005xrrc.procE4.05D} previously noted that there are small, bright features that are found in both the arcsecond-resolution radio and soft X-ray data. One of these two knots is brighter and has a steeper radio spectrum while the other is fainter with a flatter radio spectrum. The fact that \NS \ sees comparably bright X-rays from both of these knots may suggest that there may be some unknown localized physical process acting on small scales to produce the enhanced radio emission as well as the enhanced hard X-ray emission. However, the data that we are comparing are not contemporaneous and Cas A is known to vary in both the radio \citep{1996AAS...188.7403R} and in the soft X-rays \citep{2007AJ....133..147P, 2008ApJ...677L.105U}. We note that there now exists Janksy VLA radio images \citep{DeLaney:2014kj} that are contemporaneous with these \NS \ observations and that these knots have not significantly changed their brightness or morphology when compared to the older VLA image. It may be the case that the bright features in the Cas A hard X-ray images are also evolving on timescales of a few years, though this is unlikely to explain the large-scale discrepancies between the radio and the hard X-ray images.

While the $>$15 keV emission is clearly dominated by several central knots and the outer filaments, there is also diffuse emission that appears to permeate the remnant. While the central knots may be localized regions of enhanced magnetic field, density variations, or particle acceleration, the source of the diffuse central emission at energies $>$15 keV is unclear. It is possible that this is in fact tenuous diffuse emission from the forward shock of the supernova remnant seen in projection, or that there is a population of unresolved and relatively dim knots in the interior of the remnant. One additional possibility is that some electrons that happen to be in regions of lower mean magnetic field strength and thus are longer lived. Electrons radiating at $\approx$15 keV in a magnetic field of 20 $\mu$G or less could survive for more than the 330 year lifetime of the remnant and so could be responsible for this emission.

The hard X-ray morphology may also have implications for the interpretation of the TeV and GeV emission. The centroids of the TeV emission may be spatially consistent with the bright western knots of synchrotron emission observed by \NS, while the {\it Fermi} GeV centroid has no obvious counterpart in the \NS \ band. This might be expected if the TeV emission is leptonic in nature (i.e. inverse Compton scattering of CMB photons), which requires the presence of relativistic electrons which would also produce synchrotron emission in the \NS \ bandpass. Likewise, if the GeV emission is instead produced via hadronic emission mechanisms (that is, $\pi^{0}$ decay) then we might not expect to observe a counterpart in the \NS \ band, since the $\pi^{0}$ emission is produced by ions with much lower energies than the
synchrotron-emitting electrons observed by \NS. This would be consistent with broadband spectroscopic analyses (e.g.  \citealt{2010ApJ...720...20A} and \citealt{Saha:2014ep}), which suggests the need for both hadronic and leptonic emission mechanisms. It's also suggestive that the GeV centroid falls near enhanced emission observed at 24 and 70 microns \citep{2004ApJS..154..290H} as well as near bright optical features in this region, both of which imply the presence of dense material and further supports a hadronic emission mechanism in that region. However, as the difference in the centroid regions is not strictly statistically significant (recall that all of the centroids are consistent the 90\% level), we cannot make firm claims about separating the emission mechanisms without further improvements in the angular resolution of future GeV and TeV instruments. When those data are eventually available, combining the GeV/TeV data with these \NS \ observations should prove fruitful.

\subsection{Curvature in the continuum spectrum}

%

For the following discussion, we presume that the emission in the \NS \ band is
synchrotron radiation from electrons with 10 -- 100 TeV energies,
produced by diffusive shock acceleration.

The shape of the spectrum carries information about the source electron spectrum.
For synchrotron emission (which is the only viable mechanism for the emission at the forward shock) 
from a power-law electron spectrum with an exponential cutoff, the resulting spectrum drops roughly as
$\exp^{(-\sqrt{\nu / \nu_{\rm rolloff}})}$, where $h \nu_{\rm rolloff}$ is 1.9 times the photon
  energy at which electrons with $E_m$ radiate the peak of their
  synchrotron spectrum (e.g. \citealt{Pacholczyk:1969uv}).
 However, more elaborate inhomogeneous
  models of shell supernova remnants can predict significantly
  different spectral shapes (e.g. \citealt{Reynolds:1998jb}). In the case of the XSPEC {\tt srcut} model,
  the model produces a smooth continuum over many decades in energy, from
  the radio to the X-ray band, with a characteristic frequency at which the spectrum gently rolls over. These frequencies are in the range $10^{16}$ -- $10^{18}$ Hz. This is sufficiently slow that over a relatively narrow bandpass in the hard X-rays the resulting spectrum can be reasonably fit by a power-law (i.e. 15--50 keV), though we do expect some steepening of the spectrum from the energy band previous sampled by \CHAN \ to the \NS \ band.  
  
At the exterior of the remnant we find that the 15--50 keV power-law model under predicts the emission at low energies and therefore does not explicitly require curvature in the spectrum. This is not surprising, as we expect there to be some ``PSF bleed" of the central regions of of the remnant that provides an additional soft ($<$10 keV) component on top of the non-thermal continuum. However, in this region the non-thermal continuum spectrum as observed by \CHAN \ is well-fit by a power-law spectrum with a substantially harder photon index (e.g., $\Gamma \sim-2.3$ for the northeast filaments from \citealt{2009ApJ...697..535P}). Given that the \CHAN \ data are uncontaminated by neighboring thermal emission, we can then assume that the power-law index steepens from $\Gamma \sim-2.3$ to $\Gamma \sim-3.05$ from the 4--6 keV band to the 15--50 keV band, with no further curvature required by our data across the 15--50 keV band.

In the central emission regions the \NS \ data alone demonstrates that the softer power-law fit in the 15--50 keV band over-predicts the 3--15 keV observed emission. We actually expect more contamination by thermal plasma(s) in the central extraction regions. This implies that, if anything, there is more curvature in the central knots than in the exterior filaments.

We can naively apply the {\tt srcut} model to analyze the data to compare the \NS \ 15--50 keV
results to the 4--6 keV results from \CHAN. The {\tt srcut} model predicts a gradually steepening spectrum above
$h\nu_{\rm rolloff}$.  We calculated a custom {\tt srcut}
model for the steep radio spectral index of Cas A, $\alpha = -0.77$,
and calculated the slopes over frequency ranges of a factor of
$10^{0.5}$, roughly the photon energy range from 15--50 keV.  Then
  the measured values of $\Gamma$ uniquely predict $h \nu_{\rm
    rolloff}$ for a simple single {\tt srcut} component.  We find that the
  central-knot value of $\Gamma = -3.35$ requires $h \nu_{\rm rolloff}
  = 1.3$ keV, while the outer-filament value of $\Gamma = -3.06$
  requires $h \nu_{\rm rolloff} = 2.3$ keV.  We can perform the same
  operation for the power-law fits in the range of 4.2--6 keV done
  with {\sl Chandra} data \citep{Patnaude:2011hv}\footnote{While they find a
  significant steepening between 2000 and 2011, the implied change
  between the 2011 fits and the epochs of the {\sl NuSTAR}
  observations is considerably smaller than the errors and we shall ignore
  it.} From their 2011 data, they find $\Gamma$ of  $-2.85$ and
  $-2.56$ for a sample of exterior filaments and for a region
  similar to the \NS \ central knots, respectively, implying $h \nu_{\rm
    rolloff}$ values of 0.44 and 1.1 keV.  Thus, while our 15--50
  keV power-law indices are steeper than their 4.2--6 keV values,
  they do not steepen as much as one would expect from a single
  power-law electron spectrum with an exponential cutoff.

We conclude that while softening of the spectrum is evident, it requires either
  electron distributions with a range of cutoff energies, as might
  naturally be expected from integrating over multiple shock regions, or a modification
  of shock acceleration physics that produces a
  more gradual cutoff in the electron distribution than exponential.
  Since one careful calculation of this cutoff  \citep{Zirakashvili:2007ke} yields a distribution
  which at the shock is actually exponential in the square of electron
  energy, i.e., a much steeper cutoff, this
  latter possibility seems less likely.  We infer that even over the
  spatially localized regions of the forward blast wave and of the
  central knots, conditions must vary sufficiently to provide a range
  of electron cutoff energies whose superposition gives us the very
  gradual steepening we observe. Future {\sl NuSTAR} observations of
  the historical Type Ia SNRs may show if they share this
  property. 
  
We do urge caution in using the $h\nu_{\rm rolloff}$ frequency reported above as a proxy for the maximum
energy of the emitting electrons. Doing so requires some knowledge of the radio-spectral index (assumed to 
be 0.77 above), which is known to vary across the remnant \citep{Anderson:1996bk} as well as spatially resolved knowledge
of the radio 1 GHz flux. The latter measurement often suffers from a lack of dynamic range in Cas A (i.e. the northeast rim is undetected at radio wavelengths but bright in synchrotron X-rays). This introduces enough degeneracies in the {\tt srcut} model fits that we do not include fits here with all of the parameters allowed to vary. However, the above analysis is entirely self-consistent and our result is relatively independent of the particular parameters used in {\tt srcut}.

\subsection{Comparison with spatially-integrated measurements}

No previous instrument could spatially resolve the different emission regions in Cas A above 10 keV.  However by considering the power law spectral index integrated over the remnant it could be possible to confirm our finding that the softer inner region dominates the total flux. The integrated emission detected by previous hard X-ray non-imaging instruments (e.g., \citealt{1996AAS..120C.357T, 1997ApJ...487L..97A,Renaud:2006gra,Vink:2000bu,Maeda:2009vla}) extends to high energy ($>$100 keV), so that, in principle, the spectral index could be well-constrained. Table \ref{tab:gammas} compiles the results of these previous measurements of the integrated spectrum of Cas A. It is clear different measurements disagree at levels greater than the formal statistical errors. This is likely due to systematics associated with the background-dominated measurements or the difference in energy bands in which the different instruments are sensitive. What is clear is that previous spatially-integrated measurements are not in agreement with one another and do not have statistical or systematic precision sufficient to confirm or rule out {\it NuSTAR's} finding that the softer central emission dominates the integrated flux.


\begin{table}
  \caption{Comparison with previous Power-law Indices}
  \label{tab:gammas}
  \begin{center}
   \leavevmode
    \begin{tabular}{lcc} \hline \hline              
  Observatory (Band)    & 	$\Gamma$    & Notes \\ \hline 
{\it CGRO} (40--120 keV)$^{i}$ & $-3.06 \pm 0.41$ & 1 \\
{\it Beppo-SAX} (12--300 keV)$^{ii}$ & $-3.3 \pm 0.05$  & 2  \\
{\it Beppo-SAX}  (30--100 keV)$^{ii}$ & $-3.1 \pm 0.4$ & 2   \\
{\it\it RXTE} (20--100 keV)$^{iii}$ & $-3.125 \pm 0.05$ & 1  \\
{\it INTEGRAL} (21--120 keV)$^{iv}$ & $-3.3 \pm 0.1$ & 3 \\ 
{\it Suzaku} (3.4--40 keV)$^{v}$ &  $-3.06 \pm 0.05$ & 2 \\
{\bf NS Central Knots} (15--50 keV) &  $-3.35 \pm 0.06$ & 2 \\
{\bf NS Exterior Filaments} (15--50 keV) &  $-3.06 \pm 0.06$ & 2 
    \end{tabular}
    \end{center}
    Notes on error estimates:\\1: Unstated, assumed to be 1-$\sigma$; 2: 90\% confidence intervals. \\ 3: 1-$\sigma$. \\
    References: \\$i$: \cite{1996AAS..120C.357T} $ii$: \cite{Vink:2001bfa} \\ $iii$: \cite{Rothschild:2003ev} \\ $iv$: \cite{Renaud:2006gra}  $v$:  \cite{Maeda:2009vla} 
\end{table}


\section{Conclusions}

We have shown that the hard X-ray emission from Cas A up to 50 keV resolves into two main populations: fainter outer filaments and bright central knots which dominate the emission above 15 keV. These two populations show different unbroken power-law spectra over the 15--50 keV \NS \ band, with the central bright knots having a significantly softer spectra than the dim outer rims. We view this as evidence for two distinct populations of electrons responsible for the exterior and central emission and therefore argue that the central knots are in fact located in the interior of the remnant rather than at the forward shock and seen in projection.

The origin of the population of energetic electrons in the interior of the remnant remains a mystery, especially as the morphology above 15 keV does not appear to follow that of any other waveband. Some of the bright central knots appear to be spatially coincident with regions known to show rapid ($\sim$yearly) variability in both soft X-rays and the radio, which may evidence for active particle acceleration in the interior of Cas A or for significant changes in the physical environment in the center of the remnant. Future \NS \ observations with a longer temporal baseline may be able to test for changes in the flux of particular regions above 15 keV.

The steepening of the non-thermal spectrum from the \CHAN \ band to the \NS \ band requires either an electron distribution that cuts off more gradually than an exponential. While this could be due to some modification in the shock acceleration physics, we instead conclude that conditions in the forward shock blast wave produce a range of cutoff energies in the electron spectrum even on small spatial scales.

We have shown that imaging in the \NS \ band can be useful for interpreting results from the GeV/TeV energy bands. The association of the centroid of the TeV emission with a region bright in the \NS \ band and the lack of any such association with the centroid of the GeV emission can be naturally explained by a leptonic emission mechanism for the former and a hadronic emission mechanism for the latter. The interpretation that there are both hardonic and leptonic emission mechanisms at work in Cas A is consistent with broad-band spectrum fitting and may bear out as the measurements in the GeV and TeV bands improve.\\

B.G. thanks Una Hwang for the \CHAN \ 4--6 keV band image. This work was supported under NASA contract NNG08FD60C and made use of data from the \NS \ mission, a project led by the California Institute of Technology, managed by the Jet Propulsion Laboratory, and funded by NASA. We thank the \NS \ Operations, Software and Calibration teams for support with the execution and analysis of these observations. This research has made use of the \NS \ Data Analysis Software (NuSTARDAS), jointly developed by the ASI Science Data Center (ASDC, Italy) and the California Institute of Technology (USA). \\

{\it Facilities}: \NS, \CHAN, VLA, {\it Fermi}, VERITAS, HESS, 


\pagebreak

\end{document}